# Emergent Ferromagnetism at LaFeO$_3$/SrTiO$_3$ Interface Arising from Strain-induced Spin-State Transition

Menglin Zhu[1,‡], Joseph Lanier[2,‡], Sevim Polat Genlik[1,‡], Jose G. Flores[2], Victor da Cruz Pinha Barbosa[3], Mohit Randeria[2], Patrick M. Woodward[3], Maryam Ghazisaeidi[1], Fengyuan Yang[2], Jinwoo Hwang[1,*]

[1]Department of Materials Science and Engineering, The Ohio State University, Columbus, OH, 43210

[2]Department of Physics, The Ohio State University, Columbus, OH, 43210

[3]Department of Chemistry and Biochemistry, The Ohio State University, Columbus, OH, 43210

‡ Equal contribution

Corresponding authors: * hwang.458@osu.edu

**Creating new interfacial magnetic states with desired functionalities is attractive for fundamental studies and spintronics applications. The emergence of interfacial magnetic phases demands the fabrication of pristine interfaces and the characterization and understanding of atomic structure as well as electronic, magnetic, and orbital degrees of freedom at the interface. Here, we report a novel interfacial insulating ferromagnetic order in antiferromagnetic LaFeO$_3$ grown on SrTiO$_3$, characterized by a combination of electron microscopy and spectroscopy, magnetometry, and density functional theory. The epitaxial strain drives a spin-state disproportionation in the interfacial layer of LaFeO$_3$, which leads to a checkerboard arrangement of low- and high-spin Fe$^{3+}$ ions inside smaller and larger FeO$_6$ octahedra, respectively. Ferromagnetism at the interface arises from superexchange interactions between the low- and high-spin Fe$^{3+}$. The detailed understanding of creation of emergent magnetism illustrates the potential of designing and controlling orbital degrees of freedom at the interface to realize novel phases and functionalities for future spin-electronic applications.**

# 1 Introduction

Innovative manipulations of magnetic states are at the core of the growing fields of spintronics, topology, and quantum information science. Among the various stimuli that can control magnetism, interface engineering is of particular interest because fundamental interactions that govern magnetism, such as the correlation between electronic spin and orbital degrees of freedom, are inherently altered at interfaces (*1–12*). Thus, interfacing two materials not only can modify the bulk magnetism (*2, 3*) but also create novel ground states (emergent phenomena) that are absent from adjoined constituents (*1, 4, 7*). Beyond exploring new material systems, understanding the origin of emergent magnetism at the atomistic scale is crucial for designing and modeling spintronic devices.

Due to its structural and chemical tunability, the interface between different $ABO_3$ perovskites is a rich arena for discovering emergent magnetic states (*13–15*). Advances in growth techniques have enabled the creation of ultra-high-quality films down to a single unit cell thickness, providing a platform to understand how electronic, orbital, and structural modifications at interfaces alter magnetism at the atomic scale. In bulk form, the magnetic properties of perovskites are sensitive to the crystal structure. For example, subtle structural changes, such as rotations of $BO_6$ octahedra, can in some cases switch the magnetic ground state from ferromagnetic (FM) to antiferromagnetic (AFM)(*3, 4*). In thin films and heterostructures, epitaxial strain can be used not only to control octahedral tilting, but also to induce changes in metal-oxygen bond lengths that impact the crystal field splitting (*16–18*). This can lead to changes in *d*-orbital occupancy that dictate the exchange interactions between neighboring transition metal ions, potentially lead to spin-state transition and new magnetic ordering. In these ways, a structural mismatch at the interface between two perovskites provides a convenient way to tailor the properties of the materials on either side of the interface.

In this work, we present evidence that reveals an insulating FM state at the interface between AFM $LaFeO_3$ (LFO) and a diamagnetic $TiO_2$-terminated $SrTiO_3$ (STO) substrate. Further, a combination of scanning transmission electron microscopy (STEM), electron energy loss spectroscopy (EELS), and density functional theory (DFT) reveals a checkerboard arrangement of smaller and larger $FeO_6$ octahedra within a few unit cells of the interface. This arrangement, driven by epitaxial strain and the geometric constraints of octahedral connectivity, increases the crystal field splitting of $3d$-based orbitals on $Fe^{3+}$ in the smaller octahedra, leading to a spin-state transition from a high spin state (HSS) to a low spin state (LSS). While charge disproportionation driven by Coulomb interactions is well-known in transition metal oxides (*19–21*), the strain-induced spin disproportionation here is quite unusual. The different orbital occupancies in the smaller/larger octahedra with LSS/HSS lead to FM superexchange, as distinct from the AFM superexchange prevalent among HSS $Fe^{3+}$ ions in the bulk. This provides a natural explanation for the interfacial FM state observed, an uncommon occurrence in insulating oxides.

## 2  Results and Discussion

### 2.1  Structural Characterization

LFO films of varying thicknesses (3, 5.5, 15 and 33 nm, denoted as LFO$_{t\text{-nm}}$) are epitaxially grown on $TiO_2$-terminated STO using off-axis magneton sputtering (details in Method). Triple-axis X-ray diffraction (XRD, Fig. 1A) reveals clear Laue oscillations for all films, including LFO$_{3nm}$, indicating highly coherent single crystalline ordering. Rocking curve of the LFO$_{33nm}$ (002) peak (Fig. 1B) exhibits a full-width-at-half-maximum (FWHM) of 0.0058°, corroborating the highly uniform crystalline ordering of the LFO films (more details in SI).

Bulk LFO possesses distorted perovskite structure with *Pbnm* space group as shown in Fig. 1C ($a$ = 5.557 Å, $b$ = 5.5652 Å and $c$ = 7.8542 Å). The structural distortions involve rotations of the $FeO_6$ octahedron following the $a^-a^-c^+$ Glazer notation (*22*), accompanied by $La^{3+}$ displacements. Magnetically, $Fe^{3+}$ ions are antiferromagnetically coupled to their nearest neighbors via superexchange interactions through the oxygen $2p$ orbitals, resulting in a G-type AFM order along the $a$-axis of the *Pbnm* cell (*23*), as indicated by arrows in Fig. 1C. Deviation of the Fe–O–Fe bond angles from 180° induces a slight spin canting along $c$-axis due to Dzyaloshinkii-Moriya interactions, leading to a small net moment (~0.01 μB/Fe). The STO substrate is a diamagnetic band insulator with an ideal cubic perovskite structure ($Pm\bar{3}m$ space group with $a$ = 3.905 Å). The pseudocubic (pc) lattice parameter for LFO is $a_{pc}$ = 3.930 Å, leading to approximately 0.6% compressive strain.

Cross-sectional high-angle annular dark field (HAADF) image from [100]$_{PC}$ in Fig. 2A displays ordered atomic columns without misfit dislocations, confirming coherent growth. Energy dispersive X-rays (EDX) mapping (*Fig. 2*A inset and 2B) reveals a small degree of intermixing of Fe/Ti and La/Sr signals within the first unit cell at the interface, primarily attributed to STO surface terraces (±1 unit cell) and broadening of EDX signals. A closer examination of the interface in Fig. 2C shows a distinct and abrupt contrast change of atomic columns at the first unit cell (indicated with the arrow). As the intensity of the HAADF image is proportional to the atomic number, this observation confirms the $TiO_2$-terminated surface and chemically abrupt interface (*24*, *25*).

The STEM images in Fig. 2 unveil notable structural variations in LFO films near the interface. Specifically, the La columns exhibit an elliptical shape away from the interface (white ovals in **Fig. 2**C), stemming from the alternating displacement of La atoms every other unit cell along the viewing direction, a characteristic feature of $a^-a^-c^+$ tilting in bulk LFO. Near the interface, however, the La columns gradually adopt a more circular shape, resembling the Sr columns in STO, indicating a smaller magnitude of La displacements.

In addition, projected rotations of $TiO_6$/$FeO_6$ octahedra are quantified from annular bright-field (ABF) images, as shown in Fig. 2D. In line with the diminishing La displacement, the rotations gradually decrease to zero near the interface, aligning with the values observed in STO and deviating from those in bulk LFO. Analysis of ADF and ABF images from orthogonal viewing directions ($[010]_{PC}$ vs. $[100]_{PC}$) yields similar observations, as shown in Fig. S3. Further details regarding image collection and processing are available in Methods and SI.

In light of the structural variations near the interface, it is crucial to access the repercussions of suppressing octahedral tilts on the Fe–O bond lengths ($d_{Fe-O}$), which in turn control the crystal field splitting of the *d*-orbitals. In bulk LFO, $d_{Fe-O}$ ranges from 2.002–2.009 Å with a mean of 2.007 Å, and the Fe–O–Fe angles (α) are ~157. The pseudocubic lattice parameter of the orthorhombic cell, $a_{pc}$, is equivalent to the Fe–Fe distance in each direction, fluctuating narrowly from 3.927–3.932 Å with a mean of 3.930 Å. These variables are connected through the trigonometry relationship, $a_{pc} = (d_{Fe-O})\sin(\alpha/2)$. To match the lattice parameter of STO, $a_{pc}$ for LFO must be constraint to be 3.905 Å at the interface, as compared to 3.930 Å in the bulk. The 0.6% compressive strain could in principle be accommodated by further bending of the Fe–O–Fe angle, reducing $\sin(\alpha/2)$. However, to maintain octahedral connectivity with STO, the Fe–O–Fe bond angle becomes nearly linear at the interface. This accentuates the compression of $d_{Fe-O}$ necessary to maintain epitaxy with the underlying STO substrate. Assuming an isotropic response of the structure, for α = 180° and $a_{pc}$ = 3.905 Å, $d_{Fe-O}$ decreases to 1.953 Å, representing a 2.7% compressive strain with respect to bulk LFO.

*2.2  Structural Distortion Impact on Orbital Occupancy*

DFT calculations are further conducted to provide a more precise analysis of bond lengths and angles, as well as the orbital occupancy near the interface. The atomic positions obtained from STEM images along orthogonal directions are utilized to construct a 3-dimensional crystal model, providing the initial input for the interface structure (details in Method and SI). Subsequently, the structure model undergoes relaxation using DFT for further electronic and magnetic calculations. The resulting relaxed interface structure is schematically illustrated in **Fig. 3**A with arrows indicating the spin orientation. For convenience, each layer is labeled as follows: $[SrO]_n$ refers to the n-th SrO layer away from the interface, while $[STO]_n$ represents the n-th unit cell of STO away from the interface. The same convention applies to LFO.

The DFT-predicted O–Fe/Ti–O buckling angles are shown in **Fig. 3**B, along with the experimentally measured values, revealing a consistent trend of straightening within the first three unit cells near the interface, confirming the strain-induced structural perturbation. In the proximate interfacial layer ($[LFO]_1$), DFT calculations yield an average in-plane Fe-Fe distance of 3.905 Å, fulfilling the epitaxial match to the STO substrate. Analysis of the Fe-O distances in $[LFO]_1$ (Table S1) reveals a non-uniform propagation of strain across the interface. Instead, the strain is accommodated by spatially modulated impact on the $d_{Fe-O}$ within the same layer, resulting in the spontaneous formation of a checkerboard pattern of alternating large (red) and small (blue) octahedrons as shown in Fig. 3A and 3C, distinguishing them from the octahedra found further from the interface (brown). In one $FeO_6$ octahedron, all $d_{Fe-O}$ are compressed ranging from 1.968 Å to 1.910 Å with a mean of 1.931 Å, while the neighboring $FeO_6$ octahedron exhibits a range spanning from 1.948 Å to 2.032 Å, with a mean of 2.007 Å (Fig. S5). The observed disproportionate strain distribution will later be detailed as having profound implications on local magnetic moments. The average in-plane Fe-Fe distance undergoes relaxation in the $[LFO]_2$, reaching ~3.917 Å, and then further relaxes to ~3.923 Å in the $[LFO]_3$, with this value remaining consistent in the layers above. Therefore, all layer-resolved average Fe-O distances, Fe-Fe distances (Table S2) and O-Fe buckling angles (Table S3) suggest that structural perturbations are recovered within the first three unit cells.

The predicted magnetic ground state, indicated by arrows in Fig. 3A, displays behavior distinct from that expected in the bulk. For unit cells away from the interface, despite their bulk-like structure, the AFM moment aligns at 45.5±0.2° out of plane with its projection lying on the *c*-axis of *Pbnm* LFO. This deviation from the ideal bulk orientation (where the AFM axis is along *a*-axis) agrees well with previously reported X-ray magnetic linear dichroism measurements and is attributed to the broken crystal symmetry of the LFO film (*26*).

Upon approaching the interface, the AFM structure itself breaks down. The moments of Fe atoms in [LFO]$_1$, display FM coupling with alternating higher and lower magnin a checkerboard pattern (Fig. 3A), coincides with the checkerboard arrangement of large (red) and small (blue) FeO$_6$ octahedrons. One type (blue) of octahedron is compressed in all directions with an average $d_{Fe-O}$ of 1.931 Å, featuring iron with moment sitting in the O-cage. The other type (red) shows mixed distortions with an average $d_{Fe-O}$ of 2.000 Å, and iron with a moment of ~4.2 $\mu_B$. Compared to the DFT calculated average $d_{Fe-O}$ in bulk LFO (~2.000 Å), the Fe-O bonds in the compressed type is under 3.45% compressive strain, while the Fe-O bonds in the other type, referred to as "bulk-like", show a negligible deviation from the bulk value.

The observation that the FeO$_6$ octahedra have different moments can be understood by examining orbital populations (Table 1). The $Fe^{3+}$ ions sitting on the compressed site with lower magnetic moment are in LSS ($t_{2g}^5 e_g^0$); whereas the $Fe^{3+}$ ions residing in the bulk-like octahedra with larger moment are in HSS ($t_{2g}^3 e_g^2$), similar to those in bulk LFO. The $d_{Fe-O}$ also exhibits slight variations in [LFO]$_2$, where half of the octahedra have average $d_{Fe-O}$ of ~1.980 Å (see Table S1) while the other half are bulk-like. However, this slight compression does not induce a spin-state transition. Therefore, the correlation between the checkerboard ordering of HSS and LSS $Fe^{3+}$ ions does not extend beyond the first layer.

*2.3 Spin-state Disproportionation*

The presence of spin disproportion was further examined experimentally using EELS at the oxygen K-edge. An EELS spectral map was collected across the interface and data from unit cells in each layer (Fig. 4A) were averaged to obtain layer-resolved O-K edges (Fig. 4B). The spectra are color-coded based on spatial locations and will be referred to as [LFO]$_n$ or [STO]$_n$. To account for signal intermixing between [LFO]$_1$ and [STO]$_1$, the interface is labelled as [LFO/STO]$_1$. Further details about data collection and processing can be found in the Methods and SI.

The fine structure of O-K edges in Fig. 4B reveals three major peaks, corresponding to transitions from the O *1s* states to the unoccupied O *2p* states hybridized with *3d* of Ti/Fe (peak 1), *4d* or *5d* of Sr/La (peak 2), and *4s/4p* of Ti/Fe (peak 3), respectively(*27*). Fine structure of peak 1, specifically, is sensitive to the *3d* orbital occupation and, consequently, the spin states of Fe.

The oxygen cage surrounding Ti/Fe atoms breaks the degeneracy of five *3d* orbitals into three $t_{2g}$ and two $e_g$ orbitals, with the energy gap of crystal field splitting (CFS) (*16–18*). Consequently, two sub-peaks are observed under "peak 1" for [LFO]$_3$ to [LFO]$_6$, corresponding to the $t_{2g}$ and $e_g$ orbitals with $t_{2g}^3 e_g^2$ occupation (HSS $Fe^{3+}$ with 5 $\mu_B$). Interestingly, the two sub-peaks merge into a broad peak as we approach the interface at [LFO/STO]$_1$. The peak merging is not the result of signal intermixing as discussed in SI (Fig. S6). Instead, it suggests a change in the occupation of the *3d* orbitals and a spin-state disproportionation consistent with the DFT prediction of the HSS-LSS checkerboard configuration (Fig. 3A), wherein half of the $Fe^{3+}$ ions in [LFO]$_1$ undergo a transition from HSS to LSS with $t_{2g}^5 e_g^0$ configuration, while the other half retains HSS with $t_{2g}^3 e_g^2$. For the LSS $Fe^{3+}$, all 5 electrons in *3d* resides in $t_{2g}$ orbitals while the $e_g$ orbitals are empty, leading to alteration in the relative intensities and positions of the two peaks.

To validate the experimental findings from EELS, we calculated the X-ray absorption spectra (XAS) at the O K-edge for HSS ($t_{2g}^3 e_g^2$, ground state) and LSS ($t_{2g}^5 e_g^0$) $Fe^{3+}$ ions in bulk LFO (*27*). The average $d_{Fe-O}$ in the LSS-constrained structure upon relaxation is ~1.921 Å, contrasting with average $d_{Fe-O}$ of 2.000 Å observed in the HSS ground state. As a result, the Fe-O bonds in the LSS-constrained relaxed structure undergo approximately 3.95% compressive strain, elucidating the presence of LSS iron within the compressed-type interfacial octahedra of [LFO]$_1$.

DFT-calculated XAS spectra (Fig. 4C) for bulk LFO revealed distinct spectral features for the HSS and LSS $Fe^{3+}$. In HSS, peak 1 (526 to 531 eV) exhibits a pronounced splitting, which is absent in LSS. Although direct comparison of absolute peak positions and intensity is limited due to fundamental computational constraints, this peak splitting serves as a spectral fingerprint for the HSS $Fe^{3+}$. Considering that half of the $Fe^{3+}$ in [LFO]$_1$ undergo HSS-to-LSS transition, we computed the average of HSS and LSS spectra, depicted by the dashed line. This average spectrum lacks clear peak splitting and shares a similar overall shape experimental EELS spectrum from [LFO]$_1$ in Fig. 4B, confirming a partial transition between HSS and LSS.

The structure mismatch at the interface prompts the formation of a checkerboard pattern comprising compressed and bulk-like FeO$_6$ octahedra, as predicted by DFT (Fig. 3C). The compressed Fe–O bonds increase the crystal field splitting of the $t_{2g}$ and $e_g$ orbitals, destabilizing the HSS. Once the crystal field splitting exceeds the Hund's coupling energy, which favors spin alignment in orbitals, the Fe$^{3+}$ ions in the compressed octahedra undergo a transition from HSS to LSS as detailed in the SI. Notably, high-pressure studies of bulk LFO using the Mössbauer effect have found a 1:1 mixture of LSS and HSS Fe$^{3+}$ over the pressure range 30–50 GPa (*28*), similar to the observed behavior where LFO is subjected to compressive strain by epitaxy on STO.

## 2.4  Emergent Ferromagnetic State

The exchange interaction between neighboring HSS-LSS Fe$^{3+}$ at the interface differs qualitatively from that between HSS-HSS ions in bulk LFO. The Fe$^{3+}$ ($t_{2g}^3 e_g^2$) ions exhibit AFM superexchange between half-filled orbitals, ubiquitous in Mott insulators. In the checker-board arranged at the interface, however, neighboring Fe$^{3+}$ ions in $t_{2g}^3 e_g^2$ and $t_{2g}^5 e_g^0$ configurations will exhibit FM superexchange between filled/empty orbitals, stabilized by Hund's coupling, as per the Goodenough–Kanamori–Anderson rules (*21, 29*). This provides a natural explanation of the DFT result of FM alignment of spins near the interface (Fig. 3A). While the effect likely confined to only one or a few unit cells near the interface, it should be detectable using a high-sensitivity superconducting quantum interference device (SQUID) magnetometer.

Magnetic hysteresis loops measured at a temperature of 300 K after the subtraction of the linear diamagnetic background from the STO substrate are shown in Fig. 5. Notably, LFO$_{3nm}$ has a very small saturation magnetic moment of ~0.4 µemu. Although this value surpasses the sensitivity of modern SQUID magnetometer (<1 × 10$^{-8}$ emu), measuring it possess a significant challenge due to the magnetic signals from the STO substrate. Most commercial STO substrates with a 5 × 5 × 0.5 mm$^3$ size exhibit a magnetic moment of 1 to 5 µemu. This signal is most likely due to the contamination during the cutting/slicing and polishing processes, underscoring the need for meticulous elimination to ensure reliable probing of the emergent magnetic behavior of very thin LFO films, as detailed in SI.

All four LFO films in Fig. 5A exhibit clear magnetic hysteresis at 300 K, with similar hysteresis loops observed at 25 K magnitude and shape. To compare the moment of thin films with different thicknesses, the magnetic moments of each hysteresis loop are normalized by the number of Fe atoms in the films to extract the $\mu_B$/Fe values, as shown in Fig. 5C. The LFO$_{33nm}$ exhibits a saturation net moment of 0.012 $\mu_B$/Fe, similar to the weak net magnetic moment in bulk LFO (0.01 $\mu_B$/Fe) due to spin canting. As the film thickness decreases, however, the net moment increases quickly, reaching 0.017 $\mu_B$/Fe for the LFO$_{15nm}$ and over 0.04 $\mu_B$/Fe for the 5.5 and 3 nm films (Fig. 5D). This significant enhancement of net moment per Fe with decreasing film thickness indicates a novel source of magnetic moment at or near the interface, consistent with the prediction of emerging ferromagnetic order by DFT and supported by the structural and spectral analysis discussed above.

## 3    Conclusion

The experimental observations collectively suggest the emergence of a FM state at the LFO/STO interface, distinct from the nature of either constituent in the heterojunction. Combining high-resolution STEM imaging and spectral analysis with DFT, the novel phase is revealed to arises from local structural distortions and the resulting change in the crystal fields and orbital occupancies. The altered orbital occupancies further leads to HSS/LSS disproportionation that is consistent with known behavior of LFO at high isostatic pressures. The elucidation of this mechanism offers a new paradigm for tailoring magnetic properties at the atomic scale, opening doors for innovative strategies to manipulate the interfaces and engineering function materials for advanced magnetic and spintronic devices.

**Method**

- *Film Growth*

LaFeO$_3$ epitaxial thin films are grown on SrTiO$_3$ by off axis magnetron sputtering. Double side polished SrTiO$_3$(001) substrates (from MTI corporation) are first soaked in DI H$_2$O for 5 minutes at 80°C, then dipped in buffer HF for 28 seconds, and finally thoroughly rinsed in DI H$_2$O. Next, the SrTiO$_3$ substrates are annealed at 1050°C for 2 hours in air to form TiO$_2$-terminated terraces. A sputtering target was made from cold-pressed LaFeO$_3$ powder. The SrTiO$_3$ substrates are loaded into the sputtering chamber and heated to 650°C for LaFeO$_3$ epitaxial growth. For off-axis sputtering, we use a gas mixture of Ar and 3% $O_2$ with a total pressure of 12 mTorr and an RF power of 65W, which result in a deposition rate of 60 nm/hour. Four thicknesses (3, 5.5, 15, and 33 nm) of the $LaFeO_3$ films are studied in this work.

- *Scanning Transmission Electron Microscopy*

Cross-sectional STEM samples were prepared with a standard lift out technique using an FEI Helios NanoLab Dual-Beam Focused Ion Beam (FIB). The FIB lamellas were first thinned to electron transparency with 30 kV and 5 kV $Ga^{2+}$ beam, and then were nanomilled using Fischione M1040 to remove damaged layer. High-angle annular dark field (HAADF) and annular bright field (ABF) images were collected with a Thermo Fisher Scientific Probe-Corrected Themis Z S/TEM microscope operating under 300 kV. The convergence angle was 20 mrad and the inner collection angles were 82 and 10 mrad, respectively unless otherwise specified. Twenty fast-scanned images (1 μs dwell time) were acquired for each acquisition and then aligned with non-rigid registration algorithm to improve precision and signal-to-noise ratio.

- *Monochromated Electron Energy Loss Spectroscopy*

Monochromated electron energy loss spectroscopy (mono-EELS) maps were collected with Thermo Fisher Scientific uncorrected Titan3™ G2 60-300 S/TEM and Gatan K2 summit camera. The microscope was operated under 300 kV with a convergence angle of 7.5 mrad and collection angle of 14 mrad. The electron beam is monochromated to improve the intrinsic energy resolution to ~0.15 eV and the probe current is ~40 pA. The energy dispersion was 0.5 eV/channel for elemental mapping and 0.1 eV/channel for electron energy loss near edge structure (ELNES) analysis, and the dwell time was 0.1 s/pixel. All spectrums were subjected to a power-law background subtraction and no other denoise filter was applied. More details of spectrum processing and analysis can be found in Supplementary Materials.

- *Density Function Theory Calculations*

Density Functional Theory (DFT) calculations are performed with Vienna ab initio Simulation Package (VASP)(*30*) using projector augmented wave (PAW) pseudopotentials(*31*). Exchange correlations were treated by Strongly Constrained Appropriately Normed (SCAN)(*32*) meta-GGA functional to capture strong electron correlations in partially filled *d* shells. Structural optimizations and electronic structure calculations were carried out with a plane-wave cutoff energy of 375 eV and k points density of 0.2 Å$^{-1}$. Initial atomic positions were obtained from STEM image analysis and all atomic positions were fully relaxed until forces are less than 5 meV Å$^{-1}$. Full periodic boundary conditions were used for large heterostructure supercell consisting of 5 nm thick LaFeO$_3$ (13 unit cells) and 3nm thick SrTiO$_3$ (8 unit cells) layers to ensure accurate distortions and energies. Two symmetric n-type interfaces with TiO$_2$-LaO terminations were considered. Spin orbit coupling effects were considered for non-collinear magnetism. SCAN functional performed well to reproduce ground state structural and magnetic properties of bulk LFO. Lattice constants are predicted as a = 5.553 Å, b = 5.569 Å, and c = 7.892 Å. The average Fe-O distance in bulk LFO is calculated to be approximately 2.000 Å, accompanied by O-Fe buckling angles of ~158.8°. Notably, the SCAN functional accurately captured the G-type AFM ordering, with the AFM axis along the orthorhombic a-axis, representing the ground state magnetic ordering. A slight spin canting angle of approximately 0.52° was observed, resulting in a net magnetization along the c-axis of 0.027 µ$_B$/Fe.

All X-ray Absorption Spectroscopy (XAS) spectrum calculations were carried out using bulk LFO, not the heterostructure, due to the limitation of computational resources. The supercell core-hole method is employed to calculate the XAS spectrum of K-edge of oxygen ions in bulk LFO with different spin states of the Fe$^{3+}$. GW(*33*) pseudopotentials are utilized with a plane-wave cutoff energy of 425 eV. To set the occupations of d orbitals of Fe$^{3+}$ in high and low spin-state configurations, occupation matrix control(*34*) in the framework of VASP together with DFT+U corrections(*35*) is used. In this method, polaronic distortions are first obtained through relaxation under constrained occupation conditions. Subsequently, the system is re-relaxed without occupation control. To eliminate core-hole interactions with its periodic images, 2x2x2 orthorhombic *Pbnm* LFO supercell with 160 atoms is used. Additionally, we applied the Dudarev method(*36*) to DFT+U, which employs a single effective Hubbard parameter, U$_{eff}$ = U-J, where U is the Hubbard repulsion and J is the intra-atomic exchange. Both of U and J pertain to the electrons in the localized *d*-states of Fe$^{3+}$. U$_{eff}$ is set to 4.8 eV, which successfully reproduces the band gap of LFO as 2.1 eV ( E$_g^{Exp}$ = 2.07 eV)(*37*) and spin state of Fe$^{3+}$ as high spin state ($t_{2g}^3 e_g^2$).

**Author Information**

Conceptualization: P.M.W., F.Y., and J.H. Methodology: M.R., P.M.W., M.G., F.Y., and J.H. Investigation: M.Z, J.L., S.P.G., J.F., V.d.C.P.B., and M.R. Supervision: P.M.W., F.Y., and J.H. Writing–original draft: M.Z., and J.H. Writing–review and editing: M.R., P.M.W., M.G., F.Y., and J.H.

**Acknowledgement**

Authors acknowledge the support by Center for Emergent Materials (CEM), a National Science Foundation MRSEC under NSF Award Number DMR-2011876. Electron microscopy was performed at the Center for Electron Microscopy and Analysis (CEMAS) at The Ohio State University. Computational resources were provided by the Ohio Supercomputer center.**References**

1. A. Hoffmann, J. W. Seo, M. R. Fitzsimmons, H. Siegwart, J. Fompeyrine, J. P. Locquet, J. A. Dura, C. F. Majkrzak, Induced magnetic moments at a ferromagnet-antiferromagnet interface. *Phys. Rev. B - Condens. Matter Mater. Phys.* **66**, 1–4 (2002).

2. K. Kjærnes, I. Hallsteinsen, R. V. Chopdekar, M. Moreau, T. Bolstad, I.-H. Svenum, S. M. Selbach, T. Tybell, Uniaxial Néel vector control in perovskite oxide thin films by anisotropic strain engineering. *Phys. Rev. B* **103**, 224435 (2021).

3. A. J. Grutter, S. M. Disseler, E. J. Moon, D. A. Gilbert, E. Arenholz, A. Suter, T. Prokscha, Z. Salman, B. J. Kirby, S. J. May, Strain-induced competition between ferromagnetism and emergent antiferromagnetism in (Eu,Sr) MnO3. *Phys. Rev. Mater.* **2**, 094402 (2018).

4. J. S. White, M. Bator, Y. Hu, H. Luetkens, J. Stahn, S. Capelli, S. Das, M. Döbeli, T. Lippert, V. K. Malik, J. Martynczuk, A. Wokaun, M. Kenzelmann, C. Niedermayer, C. W. Schneider, Strain-induced ferromagnetism in antiferromagnetic LuMnO3 thin films. *Phys. Rev. Lett.* **111**, 037201 (2013).

5. F. Nolting, A. Scholl, J. Stöhr, J. W. Seo, J. Fompeyrine, H. Siegwart, J. P. Locquet, S. Anders, J. Lüning, E. E. Fullerton, M. F. Toney, M. R. Scheinfein, H. A. Padmore, Direct observation of the alignment of ferromagnetic spins by antiferromagnetic spins. *Nat. 2000 4056788* **405**, 767–769 (2000).12

6. S. Dong, K. Yamauchi, S. Yunoki, R. Yu, S. Liang, A. Moreo, J. M. Liu, S. Picozzi, E. Dagotto, Exchange bias driven by the dzyaloshinskii-moriya interaction and ferroelectric polarization at G-type antiferromagnetic perovskite interfaces. *Phys. Rev. Lett.* **103** (2009).

7. B. Kalisky, J. A. Bert, B. B. Klopfer, C. Bell, H. K. Sato, M. Hosoda, Y. Hikita, H. Y. Hwang, K. A. Moler, Critical thickness for ferromagnetism in LaAlO3/SrTiO3 heterostructures. *Nat. Commun. 2012 31* **3**, 1–7 (2012).

8. K. S. Takahashi, M. Kawasaki, Y. Tokura, Interface ferromagnetism in oxide superlattices of CaMnO3/CaRuO3. *Appl. Phys. Lett.* **79**, 1324 (2001).

9. A. Maignan, C. Martin, M. Hervieu, B. Raveau, Ferromagnetism and metallicity in the CaMn1−xRuxO3 perovskites: a highly inhomogeneous system. *Solid State Commun.* **117**, 377–382 (2001).

10. A. Y. Borisevich, H. J. Chang, M. Huijben, M. P. Oxley, S. Okamoto, M. K. Niranjan, J. D. Burton, E. Y. Tsymbal, Y. H. Chu, P. Yu, R. Ramesh, S. V. Kalinin, S. J. Pennycook, Suppression of octahedral tilts and associated changes in electronic properties at epitaxial oxide heterostructure interfaces. *Phys. Rev. Lett.* **105**, 087204 (2010).

11. P. Yu, J. S. Lee, S. Okamoto, M. D. Rossell, M. Huijben, C. H. Yang, Q. He, J. X. Zhang, S. Y. Yang, M. J. Lee, Q. M. Ramasse, R. Erni, Y. H. Chu, D. A. Arena, C. C. Kao, L. W. Martin, R. Ramesh, Interface ferromagnetism and orbital reconstruction in BiFeO 3- La0.7Sr0.3MnO3 heterostructures. *Phys. Rev. Lett.* **105**, 027201 (2010).

12. J. He, A. Borisevich, S. V. Kalinin, S. J. Pennycook, S. T. Pantelides, Control of octahedral tilts and magnetic properties of perovskite oxide heterostructures by substrate symmetry. *Phys. Rev. Lett.* **105** (2010).

13. R. Ramesh, D. G. Schlom, Creating emergent phenomena in oxide superlattices. *Nat. Rev. Mater.* **4**, 257–268 (2019).

14. S. Vasala, M. Karppinen, A2B′B″O6 perovskites: A review. *Prog. Solid State Chem.* **43**, 1–36 (2015).

15. A. K. Kundu, *Magnetic Perovskites* (Springer India, New Delhi, 2016;

http://link.springer.com/10.1007/978-81-322-2761-8)*Engineering Materials*.

16. S. Susarla, P. García-Fernández, C. Ophus, S. Das, P. Aguado-Puente, M. McCarter, P. Ercius, L. W. Martin, R. Ramesh, J. Junquera, Atomic scale crystal field mapping of polar vortices in oxide superlattices. *Nat. Commun.* **12**, 6273 (2021).

17. P. V. Balachandran, A. Cammarata, B. B. Nelson-Cheeseman, A. Bhattacharya, J. M. Rondinelli, Inductive crystal field control in layered metal oxides with correlated electrons. *APL Mater.* **2** (2014).

18. G. Demazeau, B. Buffat, M. Pouchard, P. Hagenmuller, Stabilization of unusual electronic configurations of transition elements in elongated six-coordinated oxygen sites of a K2NiF4 structure. *J. Solid State Chem.* **54**, 389–399 (1984).

19. M. Imada, A. Fujimori, Y. Tokura, Metal-insulator transitions. *Rev. Mod. Phys.* **70**, 1039–1263 (1998).

20. P. M. Woodward, D. E. Cox, E. Moshopoulou, A. W. Sleight, S. Morimoto, Structural studies of charge disproportionation and magnetic order in $CaFeO_3$. *Phys. Rev. B* **62**, 844–855 (2000).

21. D. I. Khomskii, *Transition Metal Compounds* (Cambridge University Press, 2014; https://www.cambridge.org/core/books/transition-metal-compounds/037907D3274F602D84CFECA02A493395).

22. C. Structure, C. Caotnta, C. Chothia, P. ; Pauling, P. Pauling, L. W. Coleman, O. E. Little, R. A. I. Bannard, D. T. Cromer, J. B. Mann, R. Guttormson, B. Robertson, R. F. Stewart, E. R. Davidson, W. T. Simpson, The classification of tilted octahedra in perovskites. *Acta Crystallogr. Sect. B* **28**, 3384–3392 (1972).

23. J. W. Seo, E. E. Fullerton, F. Nolting, A. Scholl, J. Fompeyrine, J. P. Locquet, Antiferromagnetic LaFeO3 thin films and their effect on exchange bias. *J. Phys. Condens. Matter* **20**, 264014 (2008).

24. P. Xu, W. Han, P. M. Rice, J. Jeong, M. G. Samant, K. Mohseni, H. L. Meyerheim, S.

Ostanin, I. V Maznichenko, I. Mertig, E. K. U Gross, A. Ernst, S. S. P Parkin, P. Xu, W. Han, P. M. Rice, J. Jeong, M. G. Samant, S. S. P Parkin, K. Mohseni, H. L. Meyerheim, S. Ostanin, I. Mertig, E. K. U Gross, A. Ernst, I. V Maznichenko, Reversible Formation of 2D Electron Gas at the LaFeO3/SrTiO3 Interface via Control of Oxygen Vacancies. *Adv. Mater.* **29**, 1604447 (2017).

25. S. R. Spurgeon, P. V Sushko, S. A. Chambers, R. B. Comes, Dynamic interface rearrangement in LaFeO 3 /n-SrTiO 3 heterojunctions. *Phys. Rev. Mater.* **1**, 63401 (2017).

26. J. Lü Ning, F. Nolting, A. Scholl, H. Ohldag, J. W. Seo, J. Fompeyrine, J.-P. Locquet, J. Stö Hr, Determination of the antiferromagnetic spin axis in epitaxial LaFeO 3 films by x-ray magnetic linear dichroism spectroscopy. doi: 10.1103/PhysRevB.67.214433.

27. F. Frati, M. O. J. Y. J. Y. Hunault, F. M. F. F. De Groot, Oxygen K-edge X-ray Absorption Spectra. *Chem. Rev.* **120**, 4056–4110 (2020).

28. W. M. Xu, M. P. Pasternak, O. Naaman, G. K. Rozenberg, R. D. Taylor, Pressure-induced breakdown of a correlated system: The progressive collapse of the Mott-Hubbard state in (formula presented). *Phys. Rev. B - Condens. Matter Mater. Phys.* **64** (2001).

29. V. G. Harris, P. Andalib, "Goodenough–Kanamori–Anderson Rules of Superexchange Applied to Ferrite Systems" in *Modern Ferrites* (Wiley, 2022; https://onlinelibrary.wiley.com/doi/10.1002/9781394156146.ch2), pp. 31–67.

30. G. Kresse, J. Furthmüller, Efficient iterative schemes for *ab initio* total-energy calculations using a plane-wave basis set. *Phys. Rev. B* **54**, 11169 (1996).

31. G. Kresse, D. Joubert, From ultrasoft pseudopotentials to the projector augmented-wave method. *Phys. Rev. B* **59**, 1758 (1999).

32. J. Sun, A. Ruzsinszky, J. Perdew, Strongly Constrained and Appropriately Normed Semilocal Density Functional. *Phys. Rev. Lett.* **115**, 036402 (2015).

33. L. Hedin, New Method for Calculating the One-Particle Green's Function with Application to the Electron-Gas Problem. *Phys. Rev.* **139**, A796 (1965).

34. J. P. Allen, G. W. Watson, Occupation matrix control of d- and f-electron localisations using DFT + U. *Phys. Chem. Chem. Phys.* **16**, 21016–21031 (2014).

35. V. I. Anisimov, J. Zaanen, O. K. Andersen, Band theory and Mott insulators: Hubbard U instead of Stoner I. *Phys. Rev. B* **44**, 943–954 (1991).

36. S. L. Dudarev, G. A. Botton, S. Y. Savrasov, C. J. Humphreys, A. P. Sutton, Electron-energy-loss spectra and the structural stability of nickel oxide: An LSDA+U study. *Phys. Rev. B* **57**, 1505 (1998).

37. P. Wang, Y. He, Y. Mi, J. Zhu, F. Zhang, Y. Liu, Y. Yang, M. Chen, D. Cao, Enhanced photoelectrochemical performance of LaFeO3 photocathode with Au buffer layer. *RSC Adv.* **9**, 26780–26786 (2019).

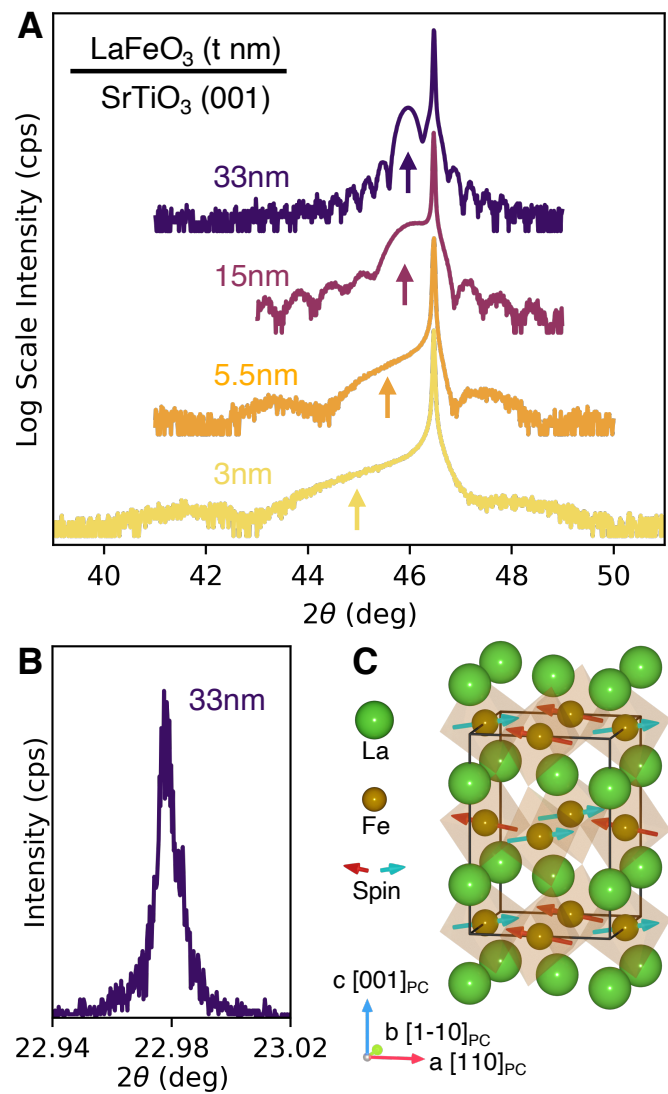

**Fig. 1** (A) $2\theta$-$\omega$ XRD scans of LaFeO$_3$ films with thicknesses of 3, 5.5, 15, and 33 nm grown on SrTiO$_3$ (001) substrates, where the arrows indicate the positions of the LaFeO$_3$ (002) peak. (B) XRD rocking curve of the LaFeO$_3$ (33 nm) film with a narrow FWHM of 0.0058°. (C) Crystal structure of LaFeO$_3$ in pseudocubic reference frame with arrows indicating orientation of Fe$^{3+}$ magnetic moments.

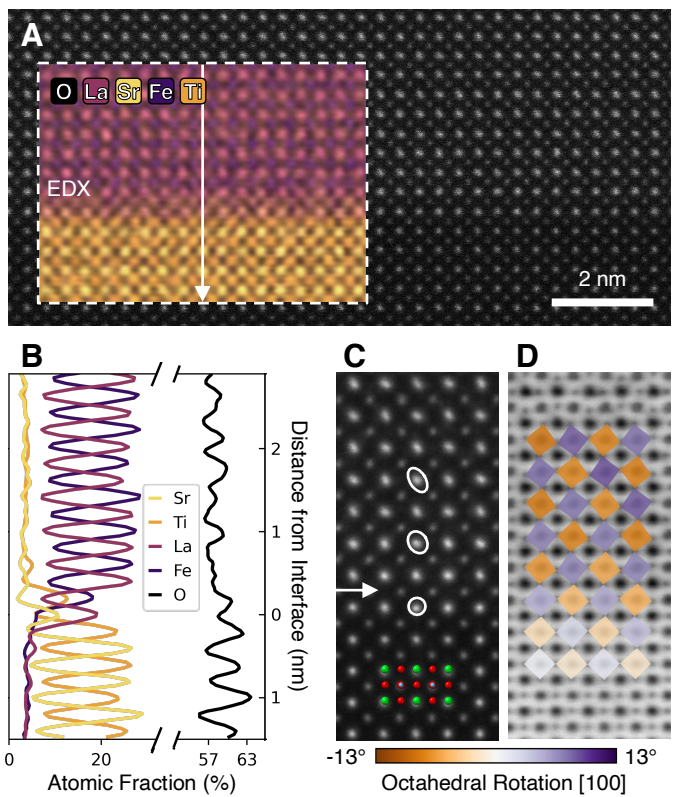

**Fig. 2** (A) Cross-sectional HAADF STEM image with overlaid EDX map illustrates a coherently strained interface without dislocations. (B) Line profile of atomic fraction along the arrow in (A) averaged over the lateral direction. (C) HAADF and (D) ABF STEM images of the interface viewed along $[100]_{pc}$. The first unit cell of $LaFeO_3$ is indicated with arrow. The white ovals in (C) highlight decreasing ellipticity of La columns near the interface resulting from smaller displacement of La due to reduced octahedral rotations. Correspondingly, the octahedral rotations are measured from ABF image in (D) and represented with squares of different colors, revealing a decreasing rotation angle near the interface.

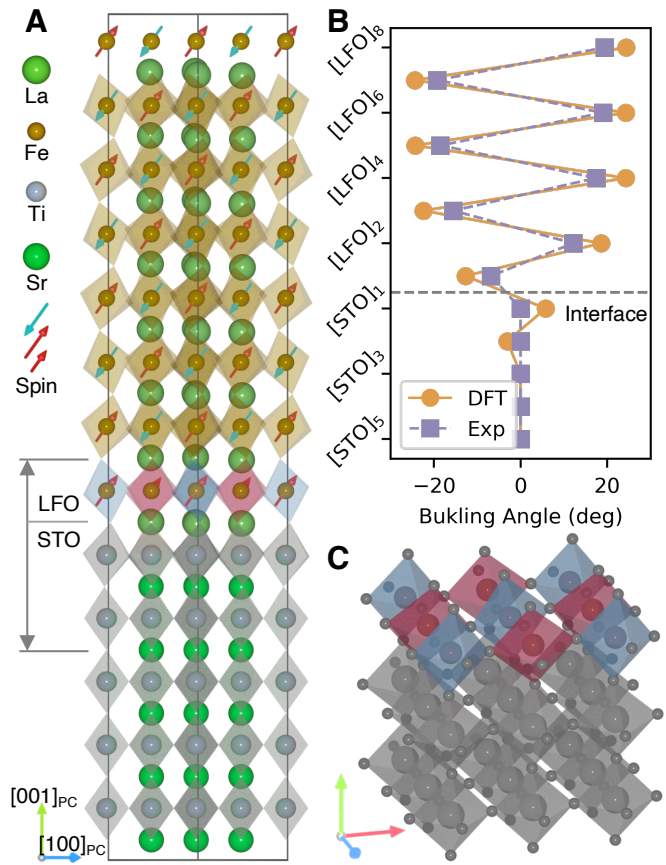

**Fig. 3** (A) DFT-relaxed structure of LaFeO$_3$/SrTiO$_3$ heterojunction is displayed from [110]$_{PC}$, with arrows indicating the magnetic moment of each Fe atom. The interface-specific compressed and bulk-like FeO$_6$ octahedra are highlighted in blue and red, respectively, distinguishing them from the rest depicted in brown. (B) The Fe-O-Fe bond buckling angle of each layer are extracted from the DFT-relaxed model and plotted alongside the experimental values. (C) Checkerboard arrangement of bigger (red) and smaller (blue) FeO$_6$ octahedra at the interface extracted from the marked regions in (A).

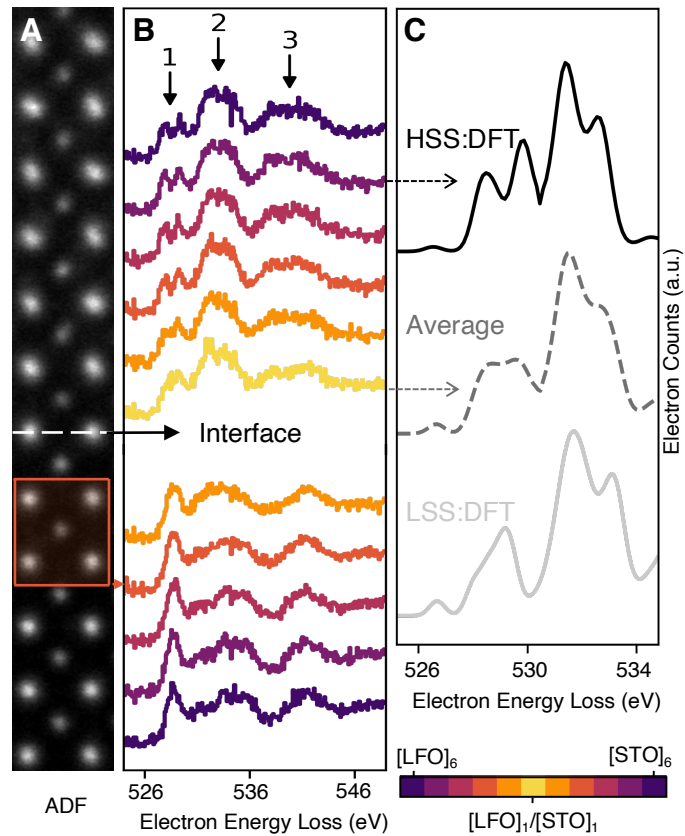

**Fig. 4** Spectra map was collected across the interface region in (A) and spectrum within each unit cell is averaged to obtain (B) layer-resolved O-K edges. (C) DFT calculated XAS spectrum of bulk LaFeO$_3$, where Fe$^{3+}$ ions are exclusively in a high spin state (HSS), low spin state (LSS). Dashed spectrum in between is the average of HSS and LSS result.

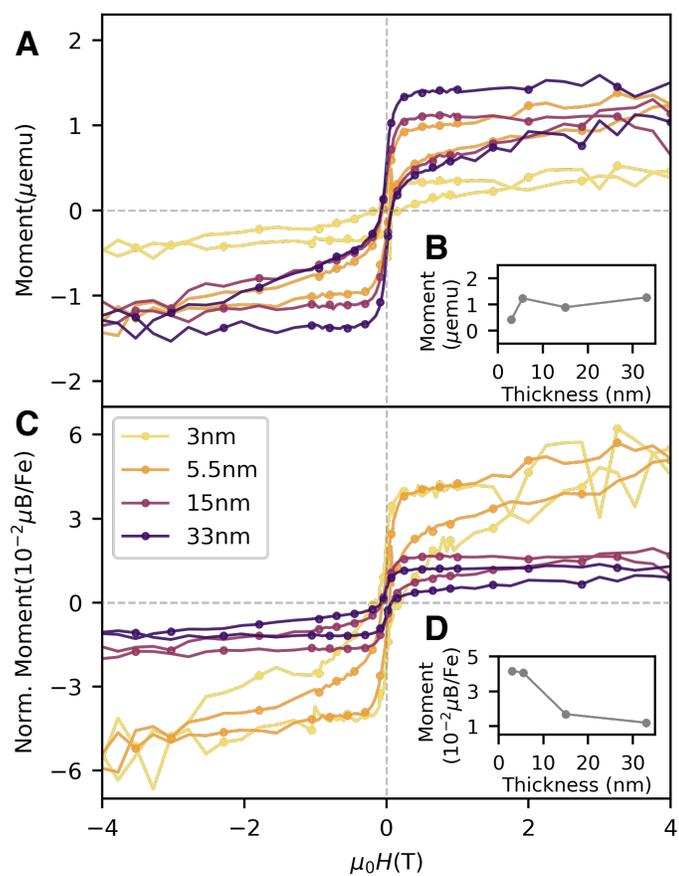

**Fig. 5** Magnetic hysteresis loops of (A) original magnetic moment and (C) normalized moment per Fe for LaFeO$_3$ epitaxial films with thicknesses of 3, 5.5, 15, and 33 nm grown on SrTiO$_3$ (001) substrates after subtraction of linear background. Inset (B) and (C): normalized magnetic moment as a function of film thickness.

**TABLE 1**. Fe $d$-orbital occupations of the two $Fe^{3+}$ sites near the interface. For the bulk-like site all five $d$-orbitals are half-occupied, where the ↓ states are filled and the ↑ states are only populated through covalency with oxygen. For the compressed site, the $d_{xz}$ and $d_{yz}$ orbitals are doubly occupied, the $d_{xy}$ singly occupied, and the $e_g$ orbitals ($d_{x2-y2}$, $d_{z2}$) are only populated through covalency with oxygen.

|  | $d_{xy}↑$ | $d_{xy}↓$ | $d_{xz}↑$ | $d_{xz}↓$ | $d_{yz}↑$ | $d_{yz}↓$ | $d_{x2-y2}↑$ | $d_{x2-y2}↓$ | $d_{z2}↑$ | $d_{z2}↓$ |
|---|---|---|---|---|---|---|---|---|---|---|
| bulk-like | 0.37 | 0.98 | 0.06 | 0.98 | 0.12 | 0.98 | 0.17 | 1.02 | 0.19 | 1.04 |
| compressed | 0.43 | 0.96 | 0.95 | 0.98 | 0.92 | 0.91 | 0.2 | 0.31 | 0.15 | 0.52 |